# Fast and wide-range wavelength tuning of a III–V/$Si_3N_4$ external-cavity laser via two-step pulsed heating

Cong Wang, Fuyi Cao, Xin Xu, Yihan Qi, Dongxin Jiang, Masataka Kobayashi, To-Fan Pan, Zhan Su, Guoen Weng, Hidefumi Akiyama, and Shaoqiang Chen

***Abstract*—Fast and wide-range wavelength switching is desirable for optical communications and photonic systems that are frequency-agile. However, thermo-optic (TO)-tuned integrated lasers often have limited switching times and tuning rates. This study demonstrates a hybrid-integrated III–V/Si3N4 external-cavity laser (ECL), combining a dual-microring Vernier filter with thermal pumping to give wide-range and fast wavelength control. The ECL provides single-mode static lasing wavelength tuning in the 1486–1614 nm range. Impulsive thermal pumping that is applied through microheaters with shorter duration and higher amplitude accelerates the switching time. A simple first-order thermal fit reproduces the measurements well, indicating that the TO-tuning dynamics are highly predictable. Consequently, two-step pulse thermal pumping is applied to the on-chip microheaters to exploit the initial quasi-linear heating transient and sustain the target wavelength at a subsequent equilibrium. The results show that 101 and 104 nm red- and blue-shift switches are achieved with quasi-linear tuning rates of 8.91 and 9.68 nm/µs, respectively. This approach provides a practical route toward fast wavelength switching in TO-tuned ECLs, potentially extending their applicability within frequency-agile systems, such as wavelength-division multiplexed transceivers.**

***Index Terms*—Integrated optics, tunable laser, fast wavelength switching, ring resonators.**

## I. INTRODUCTION

Rapid, wide-range wavelength switching is essential for telecommunication systems [1], [2], optical interconnects [3],[4], and light detection and ranging (LiDAR) equipment[5], [6]. Tunable ranges of >100 nm have been realized with hybrid-integrated external-cavity lasers (ECLs) consisting of separate gain and silicon photonics chips [7], [8], [9]. Though they possess limited tuning ranges, single microring resonators (MRRs) or Mach–Zehnder interferometers and dual (or multiple) MRR schemes can extend the effective free spectral ranges (FSRs) via the Vernier effect [10], [11], [12]. Passive platforms that are widely used include silicon (Si) [13], silicon nitride ($Si_3N_4$) [9], and silica ($SiO_2$) [14]. $Si_3N_4$ waveguides offers lower propagation loss and better stability than Si equivalents [7], [15], [16].

However, most hybrid-integrated ECLs rely on thermo-optic (TO) tuning, with switching speeds typically limited to bandwidths of a kHz-order by thermodynamics. For example, reported switching times $\Delta t_{sw}$ for $Si_3N_4$ (and corresponding wavelength ranges $\Delta\lambda_{sw}$) are as follows: 50 µs (20 nm) in Si [17], 200 µs (4 nm) [18], 100 µs (34 nm) [19], and 24.1 µs (20 nm) [20]. Among these, Guo et al. achieved the shortest switching time (2.41 µs) by using an 800 nm thick $Si_3N_4$ waveguide to strengthen mode confinement and shorten the heat diffusion path [20]. The switching time is limited by the following factors: (i) the small TO coefficient of $Si_3N_4$ [21], (ii) slower thermal convergence near the final equilibrium [13], and (iii) an increased heat diffusion path owing to the thick $SiO_2$ cladding needed to isolate the optical mode from the heater [20]. These factors hinder the broader deployment of $Si_3N_4$-based ECLs in practical settings [22], [23].

Vernier filter components have enabled the development of high-power two-step pulse heating as a simple, structure-agnostic approach [13], [24], [25]. Konoike et al. reported a tuning time of <9.4 µs for the full C band with TO heaters in a Si dual-MRRs filter [13]. However, such an approach has not yet been integrated with or applied to hybrid-integrated ECLs.

In this study, we report a hybrid-integrated $Si_3N_4$ ECL that allows for fast wavelength switching and wide-range wavelength tuning via high-power one- and two-step pulse heating. The laser is generated using dual MRRs with slightly

Received xxxxx; revised xxxxx; accepted xxxxx. Date of publication xxxxx; date of current version xxxxx.This work was supported in part by the Science and Technology Commission of Shanghai Municipality under Grant 24TS1400700, in part by the National Natural Science Foundation of China under Grant W2621019, in part by JST CREST under Grant JPMJCR2544, in part by MIC FORWARD under Grant JPMI250310003, in part by the JSPS-NSFC Joint Research Program under Grant JPJSBP120227402, in part by JSPS KAKENHI under Grants JP23K13039 and JP24K00919, in part by MEXT Q-LEAP, in part by the JAXA Space Strategy Fund under Grant JPJXSSF24MX17003, in part by the AMADA Foundation, in part by the Research Foundation for Opto-Science and Technology, in part by JST SPRING under Grant JPMJSP2108, in part by the Asahi Glass Foundation, and in part by the UTokyo-JAXA collaboration project, Japan. (Corresponding author: Cong Wang; Shaoqiang Chen.)

Cong Wang, Fuyi Cao, Yihan Qi, Masataka Kobayashi, To-Fan Pan, Hidefumi Akiyama, and Shaoqiang Chen are with the Institute for Solid State Physics, The University of Tokyo, 5-1-5 Kashiwanoha, Kashiwa 277-8581, Japan. (e-mail: wangcong@issp.u-tokyo.ac.jp; sqchen@ee.ecnu.edu.cn).
Cong Wang, Fuyi Cao, Xin Xu, Dongxin Jiang, Zhan Su, Guoen Weng, and Shaoqiang Chen are with Department of Electronic Engineering, East China Normal University, Shanghai 200241, China.

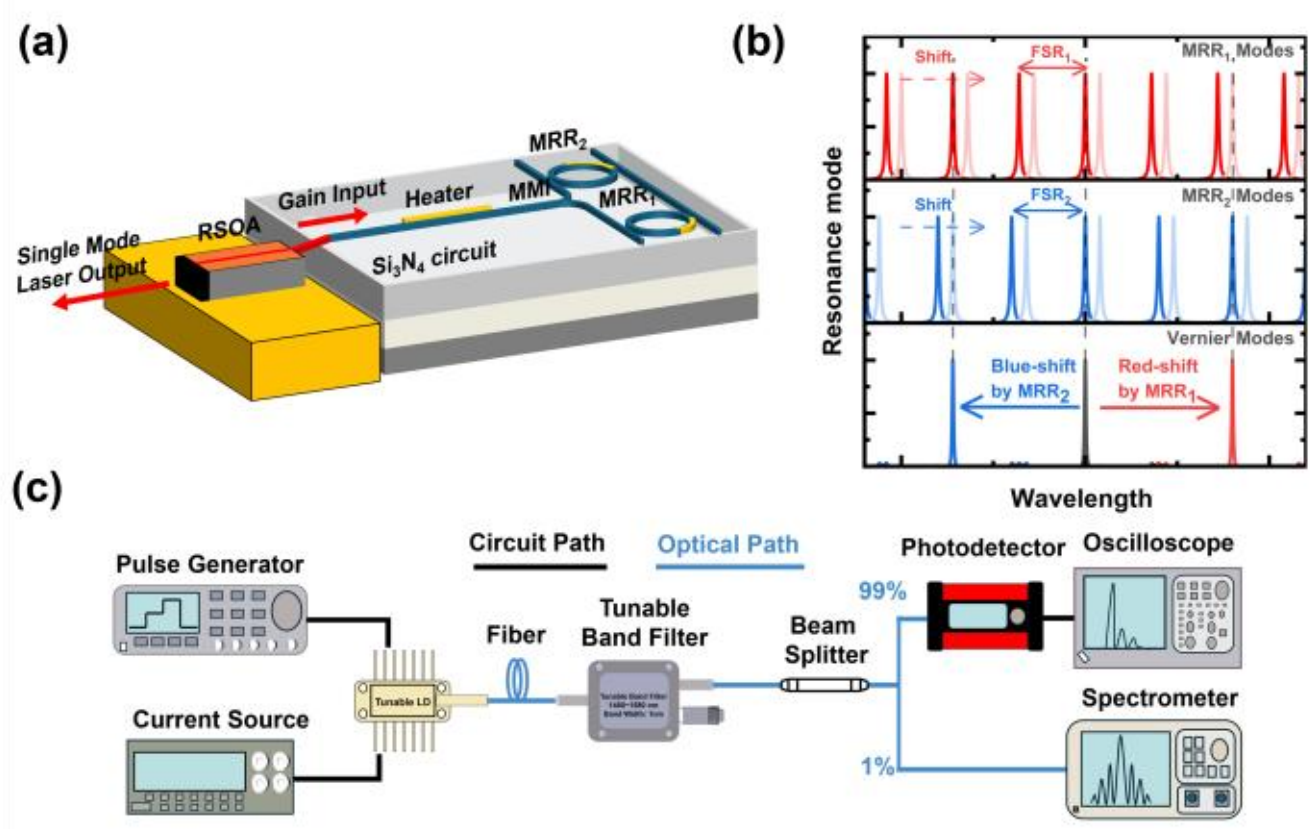

Fig. 1. (a) Schematic of the proposed III–V/$Si_3N_4$ hybrid external-cavity laser, consisting of a reflective semiconductor optical amplifier (RSOA), dual MRRs, and a multimode interference (MMI) coupler. (b) Schematic illustration of the free spectral range (FSR)-based wavelength selection and thermal tuning mechanism of the dual microring resonator (MRR) Vernier filter. The slightly different FSRs of $MRR_1$ and $MRR_2$ generate Vernier-selected modes through resonance overlap. (c) Schematic of the experimental setup for evaluating the switching time and tuning range of the laser.

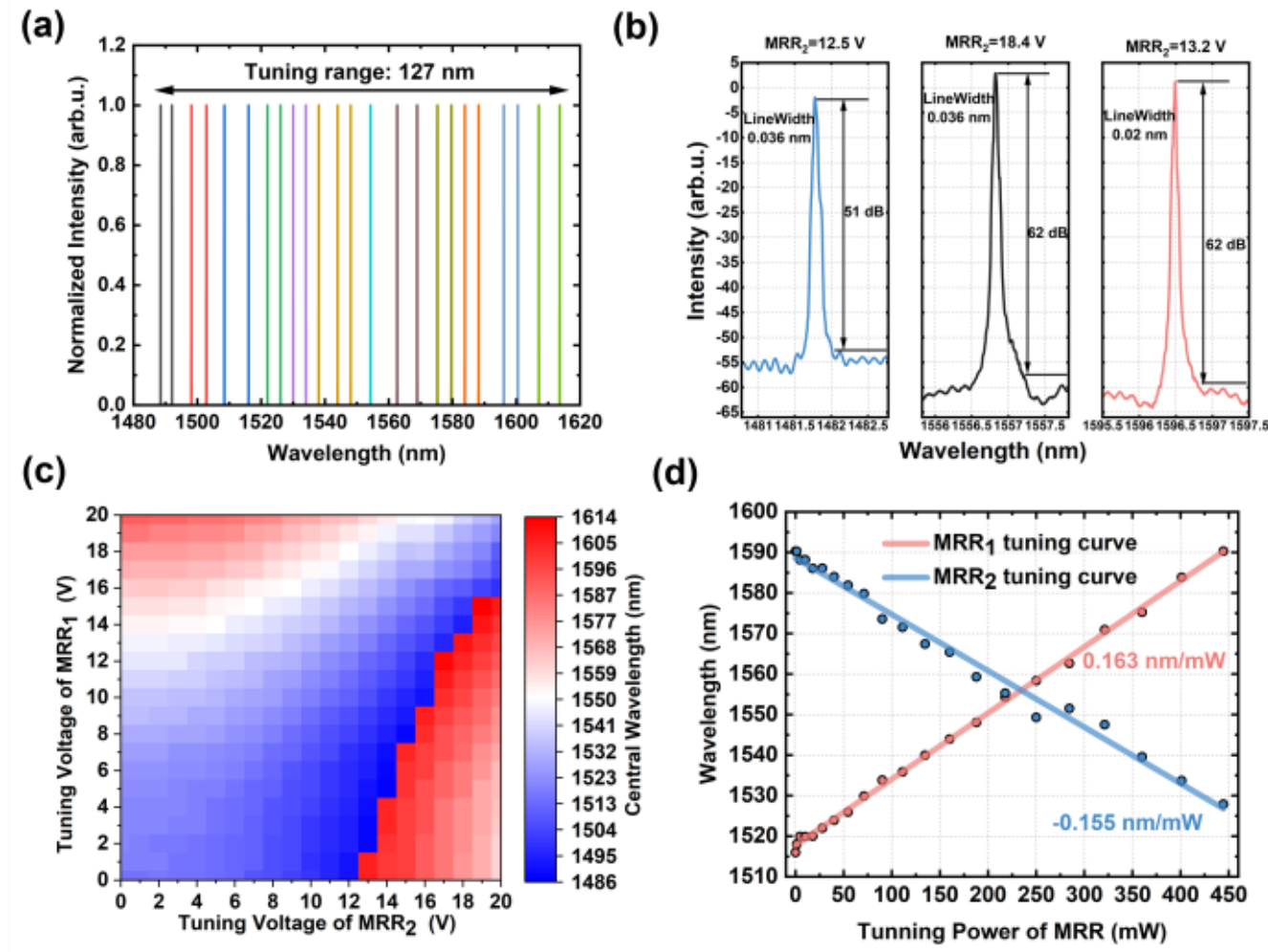

Fig. 2. Laser static output spectral performance. (a) Superimposed lasing spectra in the 1488.5–1615.9 nm range, obtained by tuning the microheaters of the dual microring resonators (MRRs). (b) Single mode lasing spectra centered on 1481.7, 1556.8, and 1596.5 nm (left to right). (c) Central wavelength mapping of single mode operation under different thermal tuning power combinations. (d) Lasing wavelengths as a function of the thermal tuning power applied to the MRR1 microheater (while the MRR2 microheater was fixed at 0 V) and the MRR2 microheater (while the MRR1 microheater was biased at 20 V).

different radii, delivering a tuning range of 127 nm. High-power two-step pulse heating enables the laser to achieve switching below 10.33 μs over a 100 nm range for red- and blue-shift tuning, corresponding to a tuning rate up of 10.36 nm/μs.

## II. DEVICE STRUCTURE, OPERATING PRINCIPLE, AND EXPERIMENTAL SETUP

The overall layout and key design choices follow those of previously reported hybrid III–V/$Si_3N_4$ ECLs [7], [20], [26]. Figure 1(a) shows the schematic configuration of the ECL, which consists of a reflective semiconductor optical amplifier (RSOA), dual MRRs, and a multimode interference coupler. The single-mode output is obtained from the left facet of the RSOA. The Si3N4 waveguide terminals integrate spot-size converters to improve mode matching and coupling efficiency.

Figure 1(b) shows the operating principle of a Vernier filter formed by two MRRs. The slightly different FSRs of $MRR_1$ and $MRR_2$ generate Vernier-selected modes through resonance overlap. Although thermo-optic heating red-shifts the resonances of each individual MRR, heating $MRR_1$ shifts the Vernier-selected wavelength to longer wavelengths, whereas heating $MRR_2$ shifts it to shorter wavelengths. Two rings with slightly different radii have resonance combs with slightly different FSRs, periodically coinciding to generate a slightly varying Vernier envelope across the spectrum. By applying different heater powers to the MRRs, the ring resonance spectra can be independently tuned, thereby shifting the Vernier envelope peak. The Vernier FSR can be expressed as:

$$FSR_{total} = \frac{FSR_1 \times FSR_2}{|FSR_1 - FSR_2|}. \qquad (1)$$

In the ECL device used in this study, the two MRRs, have FSRs of 1.96 nm ($FSR_1$) and 1.99 nm ($FSR_2$), respectively. The corresponding Vernier FSR is 130 nm, allowing for coverage for most of the S band, the entire C band, and part of the L band. The tuning is stepped by $\delta\lambda \equiv FSR_1 = 1.96$. Continuous fine tuning is enabled by including an additional heater in the ECL.

A 900 Ω microheater is integrated above each MRR to provide TO tuning to the ring. The maximum microheater DC voltage in the steady state was ~30 V, beyond which the microheater may face irreversible damage [13]. Electrical heating increases the MRR temperature, thereby changing the effective refractive index to shift the resonance to the target wavelength. A first-order TO model can describe the heating-driven wavelength response as follows:

$$\lambda_{heating}(t) = \lambda_0 + \Delta\lambda_h \left(1 - e^{-t/\tau_h}\right) \qquad (2)$$

and

$$\lambda_{cooling}(t) = \lambda_0 + \Delta\lambda_c e^{-t/\tau_c}, \qquad (3)$$

where $\lambda_0$ is the base wavelength, i.e., at zero heater power, $\Delta\lambda_h$ is the wavelength shift induced by heating, $\Delta\lambda_c$ is the wavelength shift induced by cooling, $\tau_h$ is the heating thermal time constant, and $\tau_c$ is the cooling thermal time constant. $\Delta\lambda_h$ is proportional to the electrical power $P$ dissipated in the microheater (calculated from the driving voltage) such that $\Delta\lambda_h = kP$, where $k$ is the TO tuning efficiency. In the ECL device employed in this study, heating $MRR_1$ yielded $k > 0$ and shifted the Vernier-selected resonance toward longer wavelengths (red-shift), whereas heating $MRR_2$ yielded $k < 0$ and shifted the Vernier-selected resonance toward shorter wavelengths (blue-shift). This effective blue-shift does not imply a physical blue-shift of the MRR$_2$ resonance, but results from the shift of the Vernier overlap between the two microring resonance combs.

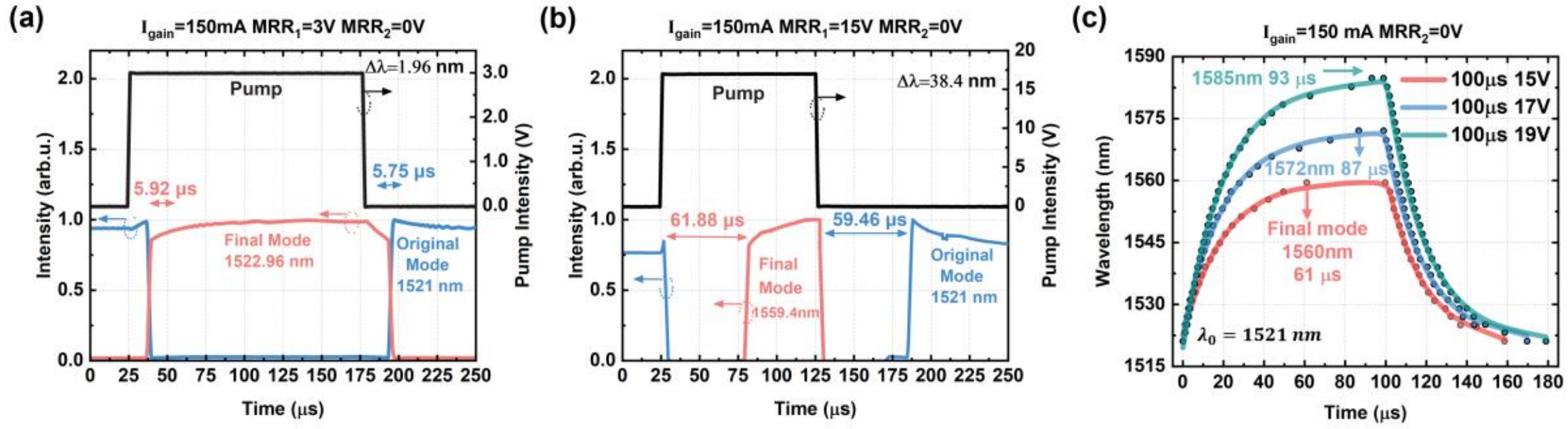

Fig. 3. Steady-state wavelength switching performance. Measured wavelength switching time of the laser for: (a) switching between adjacent Vernier-selected lasing modes from 1521.00 to 1522.96 nm, corresponding to one FSR; (b) switching between nonadjacent Vernier-selected lasing modes from 1521.00 to 1559.40 nm, corresponding to a large wavelength jump across multiple intermediate modes. (c) Measured temporal evolution of the output wavelength for different thermal pumping amplitudes.

Table I.
SWITCHING PERFORMANCE AND THERMAL TIME CONSTANTS UNDER DIFFERENT HEATER PULSE AMPLITUDES.

| Heater pulse amplitude (V) | Tuning range $\Delta\lambda_{sw}$ (nm) | Heating process | | | Cooling process | | |
|---|---|---|---|---|---|---|---|
| | | Exp. switching time $\Delta t_{sw}$ (μs) | Fit. switching time $\Delta t_{sw}$ (μs) | Fitted $\tau_h$ (μs) | Exp. switching time $\Delta t_{sw}$ (μs) | Fit. switching time $\Delta t_{sw}$ (μs) | Fitted $\tau_c$ (μs) |
| 15 | 39 | 61.05 | 66.33 | 18.08 | 58.79 | 64.50 | 17.58 |
| 17 | 51 | 86.67 | 76.89 | 19.51 | 70.74 | 72.87 | 18.49 |
| 19 | 64 | 93.10 | 89.08 | 21.34 | 80.10 | 83.52 | 20.01 |

This study defines the experimental switching time $\Delta t_{sw}$ between an initial wavelength $\lambda_i$ and a final wavelength $\lambda_f$ as the delay time using a 90% threshold, i.e., from the moment the initial mode drops to 90% of its steady-state level to the moment the final mode reaches 90% of its steady-state level. Furthermore, the switching rate can be calculated as $\Delta\lambda_{sw}/\Delta t_{sw}$ ($\Delta\lambda_{sw} \equiv |\lambda_i - \lambda_f|$.) In this study's TO model, the switching time of the heating process can be estimated by solving Eq. (2) with the limit $|\lambda_{heating}(t) - \lambda_f| < \delta\lambda/2$. Similarly, the cooling switching time can be estimated.

Figure 1(c) shows the experimental setup to evaluate the switching time and tuning range. Thermal pumping pulses at a repetition rate of 2 kHz (500 μs period) were applied to the MRR microheaters. Single-pulse heating was generated by an Agilent 81110A pulse generator (330 MHz bandwidth), whereas high-power two-step pulse heating was generated by combining the outputs of the Agilent 81110A and an HP 214B pulse generator (10 MHz bandwidth). A current source (Advantest R6240A) supplied a gain current $I_{gain}$ to the ECL, with the output sent directly or through a tunable bandpass filter (1480–1580 nm tuning range, 1 nm bandwidth) to a beam splitter. One path (99% intensity) was sent to a DC 30-GHz photodetector (Thorlabs DXM30BF) and a 20-GHz oscilloscope (Tektronix MSO 72004C), while the other (1% intensity) was sent to an optical spectrum analyzer (Yokogawa AQ6370D).

## III. LASER STATIC OUTPUT PERFORMANCE

We first evaluated the static output spectrum and tuning performance of the proposed ECL (Figure 2). The lasing wavelength was shifted across the Vernier tuning range by varying the thermal tuning power applied to the microheaters.

Figure 2(a) shows the superimposed single-mode lasing spectra in the 1488.5–1615.9 nm range, corresponding to a total tuning range of 127 nm. In the single-mode spectra, the optical signal-to-noise ratio reached 62 dB at 1556.8 nm, remained above 60 dB at 1596.5 nm, but decreased to 51 dB at 1481.7 nm due to the reduced gain of the RSOA at shorter wavelengths (Figure 2(b)). The measured 3-dB optical spectral width at the central wavelength was 4.14 GHz, as obtained from the optical spectrum analyzer (OSA). This value represents the OSA-measured spectral width, which can be affected by the spectral resolution of the OSA, and should not be interpreted as the intrinsic Lorentzian linewidth of the laser. Precise characterization of the intrinsic linewidth would require dedicated measurements, such as delayed self-heterodyne interferometry [18].

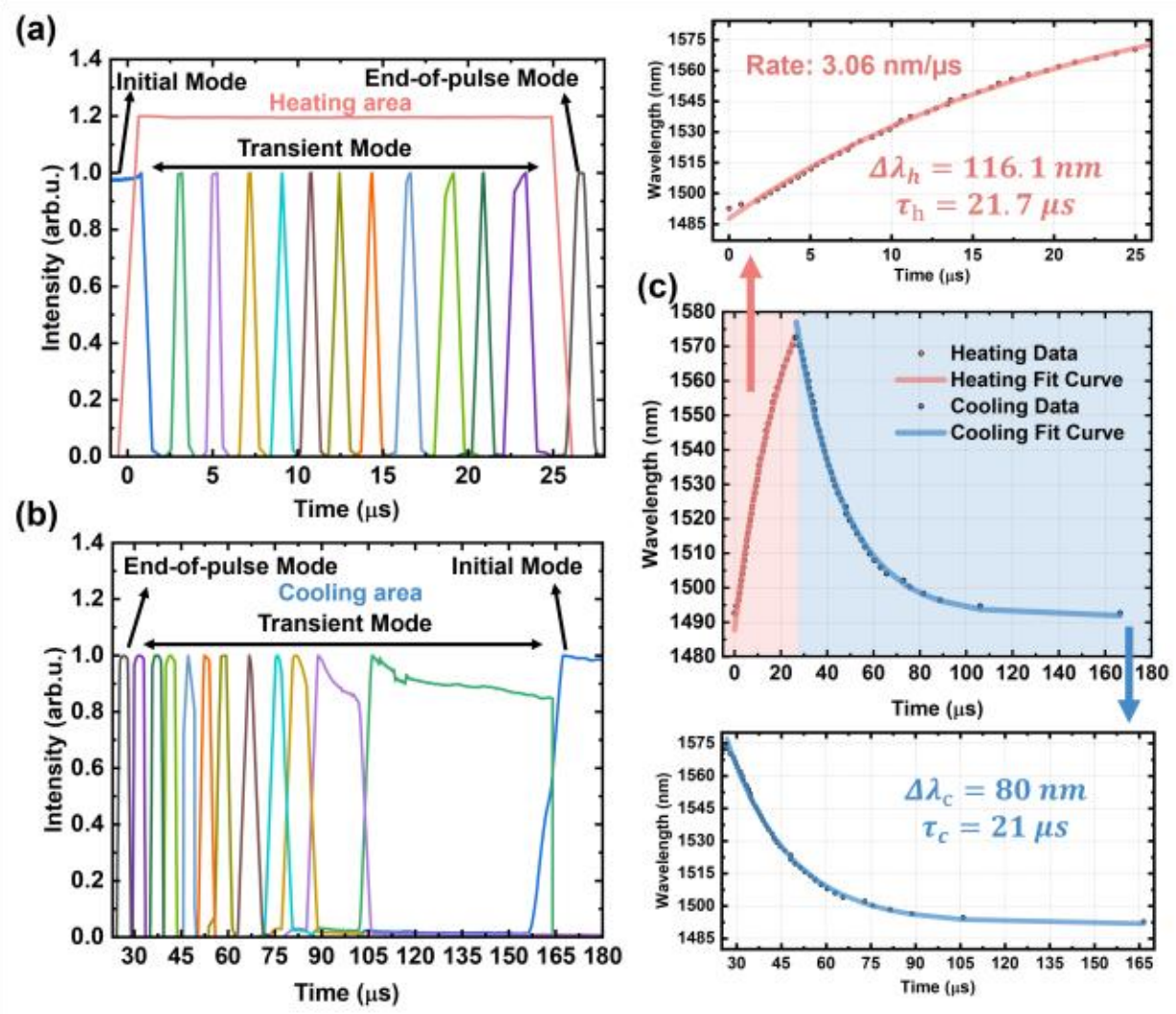

Fig. 4. Wavelength switching dynamics of the laser under short-pulse strong pumping. Transient wavelength evolution during (a) pump-induced thermal heating and (b) the subsequent cooling process. (c) Measured temporal evolution of the output wavelength during strong pumping and the subsequent cooling process.

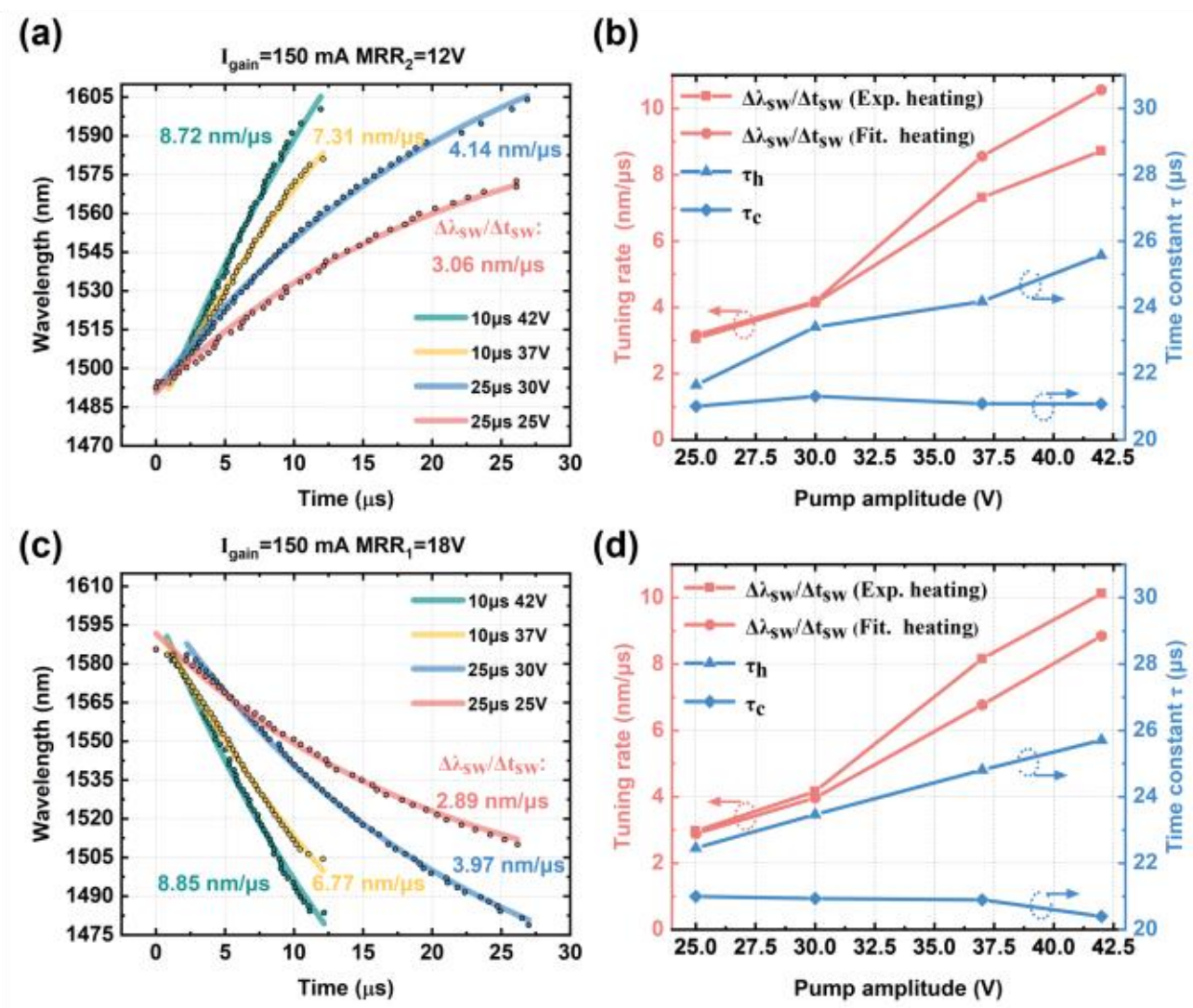

Fig. 5. Dynamics of fast wavelength switching via short pulse pumping. (a) Temporal evolution of the output wavelength and the corresponding wavelength tuning rates under different pump conditions. (b) Pump amplitude dependence of the (experimental and fitted) wavelength tuning rates and the extracted thermal time constants. (c–d) The results corresponding to (a) and (b) obtained by tuning MRR2 to induce a blue-shift.

Figure 2(c) plots the lasing wavelength as a function of the applied microheater powers on the two MRRs at a fixed gain current $I_{gain} = 150$ mA, confirming stable wavelength control across the full tuning range. To extract the individual thermal tuning efficiencies, the tuning power was varied on one MRR microheater while keeping the other fixed (Figure 2(d)). Overall, the lasing wavelength exhibits a linear dependence on the applied tuning power $P$. The extracted thermal tuning coefficients were $k_1 = 0.163$ nm/mW and $k_2 = -0.155$ nm/mW for $MRR_1$ and $MRR_2$, respectively, implying that 800–900 mW is required to shift by one Vernier FSR. These coefficients are used as fixed parameters in the subsequent fitting.

## IV. NORMAL WAVELENGTH SWITCHING

A 100 μs or 150 μs pulse drive was applied to the microheater on $MRR_1$ with a repetition rate of 2 kHz (500 μs period). The tunable bandpass filter was used to monitor the initial and final lasing modes. By varying the heating pulse amplitude, we initially switched the laser between adjacent modes and then between nonadjacent modes, and measured the corresponding switching times. Here, adjacent modes refer to neighboring Vernier-selected lasing modes separated by one MRR resonance spacing, while nonadjacent modes refer to Vernier-selected lasing modes separated by multiple resonance spacings.

A 3 V pulse ($P$ =10 mW) shifted the lasing wavelength between adjacent modes at 1521 nm and 1522.96 nm, i.e., by $\delta\lambda$ (one FSR), with a heating switching time of 5.93 μs and a cooling switching time of 5.75 μs (Figure 3(a)).

When the pulse amplitude was increased to 15 V ($P$ =250 mW), $\Delta\lambda_{sw}$ increased to 38.4 nm ($\lambda_i$ =1521 nm, $\lambda_f$ =1559.4 nm), with heating and cooling switching times ($\Delta t_{sw}$) of 61.68 μs and 59.46 μs, respectively, as shown in Fig. 3(b).

Figure 3(c) shows the temporal evolution of the output wavelength for different heater pulse amplitudes. The data were fitted by Eqs. (2) and (3). The extracted fitting parameters are summarized in Table I. Increasing the tuning power increased $\Delta\lambda_{sw}$ and extended $\Delta t_{sw}$. The heating and cooling switching times were comparable, with the heating switching time being slightly longer. The fitted model yielded comparable switching times and reproduced the same trends. Notably, the extracted thermal time constants increased with the pump amplitude for the heating and cooling processes. At a higher heating power, the temperature rise involved a larger effective thermal volume and temperature-dependent heat-transport parameters, resulting in a slower approach to equilibrium and hence a larger fitted $\tau$.

## V. FAST WAVELENGTH SWITCHING

Faster wavelength switching of the ECL was investigated using impulsive thermal pumping applied to the microheater, with a pulse width of 25 μs and a higher amplitude of 25 V across a 900-Ω load at 2 kHz. The applied voltage and the corresponding power are lower than the DC damage threshold voltage (≤30 V). Figures 4(a) and 4(b) show the corresponding time traces of the initial, transient, and final modes during the heating and cooling processes, respectively. Figure 4(c) shows the temporal evolution of the output wavelength during and after the microheater pulse, with the data fitted according to Eqs. (2) and (3). The extracted $\tau_h$ and $\tau_c$ were both ~21 μs, with $\Delta\lambda_h/\tau_h$ reaching 5.35 nm/μs. Note that a near-linear wavelength shift was observed during the heating process. The lasing wavelength switched by $\Delta\lambda_{sw}$ =81 nm ($\lambda_i$ =1491 nm, $\lambda_f$ =1572 nm) within $\Delta t_{sw}$ =26.5 μs, giving a tuning rate of $\Delta\lambda_{sw}/\Delta t_{sw}$ =3.06 nm/μs. The wavelength switched back to the initial mode during cooling in ~135 μs.

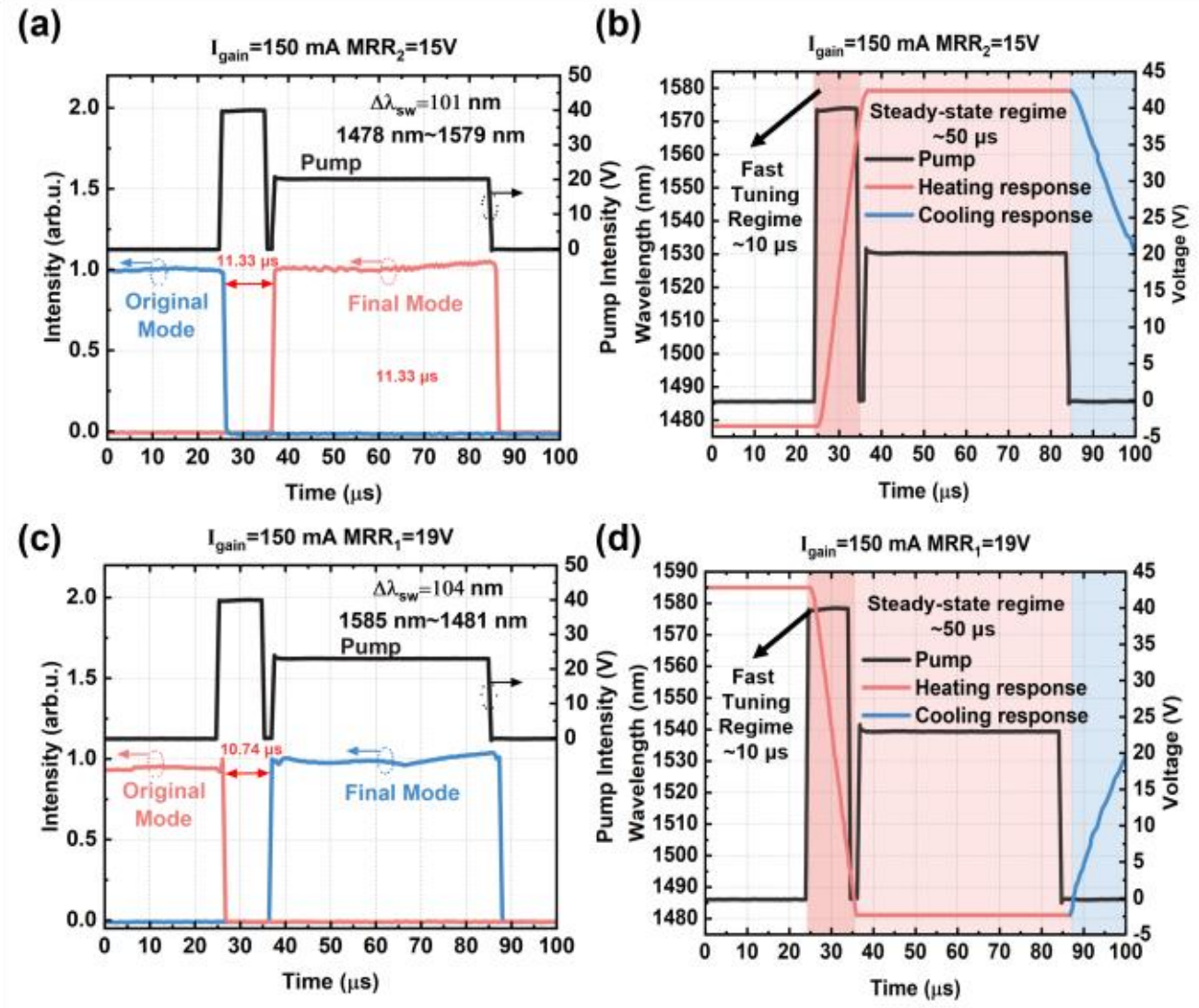

Fig. 6. (a) Fast wavelength switching enabled by two-step pulse injection. (b) Temporal evolution of the output wavelength under two-step pulse injection. (c – d) show results corresponding to (a – b) obtained when tuning MRR2 to induce a blue-shift.

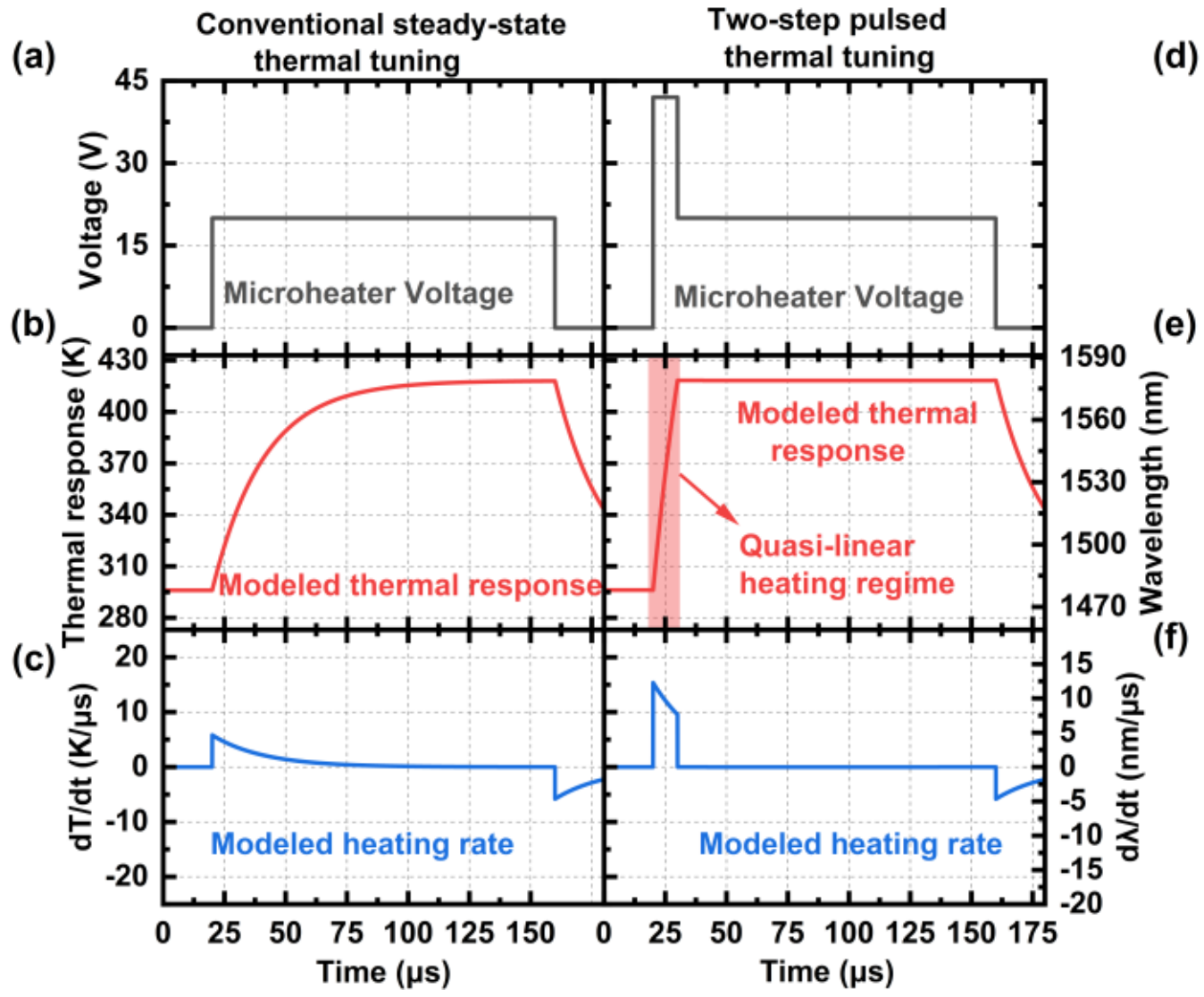

Fig. 7. Modeled thermal transient response under conventional steady-state thermal tuning and two-step pulsed thermal tuning. (a)–(c) Conventional steady-state thermal tuning: (a) pump voltage applied to the microheater, (b) modeled thermal response and wavelength evolution, and (c) modeled heating rate and wavelength tuning rate. (d)–(f) Two-step pulsed thermal tuning: (d) pump voltage applied to the microheater, (e) modeled thermal response and wavelength evolution, and (f) modeled heating rate and wavelength tuning rate.

Higher pump amplitudes are necessary to further shorten the switching time. However, the microheater drive pulse must remain below damage thresholds, which are expected to be different for DC and short-pulse drives. According to Black's equation [27], the peak power and accumulated thermal load must be limited to prevent thermal damage [13]. A similar ECL device was tested with short-pulse drives of 1 μs width; irreversible damage to the microheater occurred at 80 V peak. Therefore, to achieve high tuning rates and a wide tuning range without irreversible damage, a maximum pump voltage of 42 V was considered with a pulse width of 10 μs, i.e., 150% of the 28 V DC limit, which provides an approximate one-period wavelength shift of the Vernier filter.

Figure 5(a) shows the time-resolved output wavelength red-shifts, fitting curves, and corresponding tuning rates under different $MRR_1$ heater pulse conditions (with 12 V DC to $MRR_2$). At 37 V (10 μs), the lasing wavelength red-shifted by $\Delta\lambda_{sw}$ = 88.4 nm ($\lambda_i$ = 1492.3 nm, $\lambda_f$ =1580.7 nm) within $\Delta t_{sw}$ =12.09 μs, giving a tuning rate of 7.31 nm/μs. At 42 V (10 μs), $\Delta\lambda_{sw}$ increased to 107.7 nm ($\lambda_i$ = 1492.3 nm, $\lambda_f$ =1600 nm) with $\Delta t_{sw}$ =12.34 μs, yielding a high tuning rate of 8.72 nm/μs. Lower-amplitude and longer pulse heating resulted in lower tuning rates (3.06 nm/μs (25 V, 25 μs) and 4.14 nm/μs (30 V, 25 μs)).

The fitted wavelength tuning rates agree well with the experimental results (Figure 5(b)). Note that data for 25 V and 30 V were measured with a pulse width of 25 μs, while those for 37 V and 42 V were measured with a pulse width of 10 μs. For the 10 μs pump pulses, the fitted tuning rate was slightly higher because the experimentally measured wavelength continued to shift for ~2 μs after pump turn-off due to thermal inertia and delayed heat diffusion, effectively lowering the average rate extracted from the waveform. $\tau_h$ increased with pump amplitude, whereas $\tau_c$ remained relatively constant. This behavior could be due to short-pulse operation, where heating is localized and the temperature rise becomes more power-dependent, while subsequent cooling is primarily governed by relaxation toward the ambient thermal reservoir in the surrounding area. Figure 5(c) corresponds to Figure 5(a), but for wavelength blue-shifts induced by pulses to $MRR_2$ (with a fixed DC voltage of 18 V to $MRR_1$). Higher pulse voltages increased the switching rate. At 42 V (10 μs), the lasing wavelength blue-shifted by 102 nm ($\lambda_i = 1585$ nm, $\lambda_f = 1483$ nm) at a tuning rate of $\Delta\lambda_{sw}/\Delta t_{sw} = 8.85$ nm/μs. The corresponding pump-amplitude dependence and extracted time constants show similar trends to those observed for red-shift tuning (Figure 5(d)).

TABLE II.
PERFORMANCE COMPARISON OF SI AND $Si_3N_4$ ECLS FOR WAVELENGTH SWITCHING.

| Material platform and device | Tuning Method | Laser output Power (mW) | SMSR (dB) | Static tuning range (nm) | Switching range $\Delta\lambda_{sw}$ (nm) | Switching time $\Delta t_{sw}$ (μs) | Tuning rate $\Delta\lambda_{sw}/\Delta t_{sw}$ (nm/μs) |
|---|---|---|---|---|---|---|---|
| Si filter [13] | TO | - | 30 | 35 | 35 | 9.4 | 3.72 |
| $Si_3N_4$ ECL [28] | PZT | 11.8 | 52 | 46.16 | 21.35 | 3.86 | 5.53 |
| $Si_3N_4$ ECL [29] | PZT | 6 | 50 | 3.2 | 0.8 | 0.003 | 266 |
| Si ECL [17] | TO | 1 | 43 | 25 | 20 | 50 | 0.40 |
| $Si_3N_4$ ECL [18] | TO | 1 | 50 | 44 | 4 | 200±40 | 0.02 |
| $Si_3N_4$ ECL [19] | TO | 10 | 50 | 50 | 34 | 100 (rise) | 0.34 |
| $Si_3N_4$ ECL [20] | TO | 34 | 70 | 58.5 | 20 | 24.1 (rise) | 0.83 |
| $Si_3N_4$ ECL (this work) | TO | 10 | 60 | 127 | 101 (red shift)<br>104 (blue shift) | 11.33<br>10.74 | 8.91<br>9.68 |

## VI. TWO-STEPPED WAVELENGTH SWITCHING

Finally, the experiments were performed for two-step pulse heating, where a short high-amplitude pre-pump pulse was followed by a low-amplitude arbitrary-width main pump pulse sustaining the steady state. Pulses of 40 V (10 μs) and 20 V (50 μs) were sequentially applied to $MRR_1$ (and a fixed DC voltage of 15 V to $MRR_2$). The lasing wavelength red-shifted from 1478 nm to 1579 nm ($\Delta\lambda_{sw}$ =101 nm) with $\Delta t_{sw}$ =11.33 μs and at a tuning rate of $\Delta\lambda_{sw}/\Delta t_{sw}$ =8.91 nm/μs (Figure 6(a)).

Figure 6(b) shows the temporal evolution of the output wavelength under the two-stepped pulses. The pre-pump drove the ECL into an approximate linear fast-tuning regime, enabling the wavelength to rapidly reach the target. The subsequent main pump then established and maintained a steady-state regime. The main pump amplitude could be set to hold the wavelength at the desired value in accordance with the static tuning curve in Fig. 2(c).

Similarly, two-step pumping enabled fast blue-shift switching (Figures 6(c) and (d)). Two-stepped pulses of 40 V (10 μs) and 23 V (50 μs) were applied to $MRR_2$ (and a fixed DC voltage of 19 V to $MRR_1$). The lasing wavelength was driven from 1585 nm to 1481 nm ($\Delta\lambda_{sw} = 104$ nm) with $\Delta t_{sw}$ =10.74 μs and at a tuning rate of $\Delta\lambda_{sw}/\Delta t_{sw} = 9.68$ nm/μs.

The energy consumption associated with the two-step pulsed operation was evaluated by comparing it with the conventional 20 V thermo-optic drive. For a heater resistance of approximately 900 Ω, the peak electrical power increases from 0.44 W at 20 V to 1.96 W during the 42 V pre-pump. During the first 10 μs, the conventional 20 V drive would consume about 4.4 μJ, whereas the 42 V pre-pump consumes about 19.6 μJ. Therefore, the additional energy introduced by the high-voltage pre-pump is approximately 15.2 μJ per switching event. After the pre-pump stage, both schemes use the same 20 V hold voltage, so the additional energy cost mainly comes from this short high-power pre-pump interval. This indicates that the two-step scheme enhances the switching speed by temporarily increasing the peak power, while introducing only a limited extra energy cost for each fast-switching event.

To provide a more intuitive explanation of the physical mechanism of the two-step pulsed heating method, we compared the modeled thermal transient responses under conventional steady-state thermal tuning and two-step pulsed thermal tuning, as shown in Fig. 7. The transient response was described using a first-order thermal model with a fitted thermal time constant of approximately 21 μs. In the conventional case, a 20 V heating pulse drives the wavelength from 1478 nm toward 1579 nm through an exponential thermal response. Because the temperature rise gradually approaches its steady-state value, the wavelength transition is limited by the thermal time constant. Therefore, a relatively long time is required to reach the target wavelength.

In contrast, in the two-step pulsed scheme, a high-voltage pre-pump of 42 V is applied for 10 μs, followed by a 20 V hold pulse. The large transient heating power during the pre-pump stage produces a much higher initial heating rate and drives the heater in the early quasi-linear regime of the thermal response. As a result, the Vernier-selected wavelength can be rapidly shifted from 1478 nm to approximately 1579 nm within the 10 μs pre-pump duration. After the wavelength reaches the target channel, the lower 20 V hold pulse maintains the thermal state and stabilizes the output wavelength near the target value. This comparison illustrates that the two-step pulse does not reduce the intrinsic thermal time constant itself; instead, it accelerates the wavelength transition by using the high initial slope of the transient thermal response.

## VII. DISCUSSION AND CONCLUSION

Table II compares our results with previous reports for TO-tuned $Si_3N_4$ and Si ECLs, a piezoelectric lead zirconium titanate (PZT)-tuned $Si_3N_4$ ECLs, and a TO-tuned Si filter. Stress-optic tuning with PZTs provides rapid wavelength actuation in $Si_3N_4$ ECLs [28], [29], but the mechanical resonances limit the wavelength tuning ranges. TO-tuned Vernier Si filters displayed fast and wide switching comparable to the ECLs used in this study. However, the present work demonstrated faster and wider wavelength switching in red- and blue-shifts than other TO-tuned $Si_3N_4$ and Si ECLs.

The achievable tuning rate in our work was ultimately limited by the heat diffusion through the constraint on the maximum pulse amplitude to avoid damage to the device. The pulse amplitude and width provided straightforward control of the switching range and time in both red- and blue-shift directions. We believe that this method enables TO-tuned ECLs to be applied to enhance wavelength switching speed while maintaining a wide tuning range. More generally, this transient thermal tuning strategy can also be extended to other thermo-optically tuned Si and $Si_3N_4$ photonic devices, such as MRRs, Mach-Zehnder interferometers and programmable photonic circuits. Similar transient or pre-emphasis thermal driving strategies have also been employed in Si-based MRRs and Mach-Zehnder interferometer optical switching systems to improve the switching response of thermo-optic devices [13], [30]. By exploiting the early transient thermal response of the heater, the wavelength selection or phase tuning process can be accelerated without sacrificing the large tuning range of thermo-optic actuation.

The device was also repeatedly tested under the above pulsed-heating conditions, with each measurement session lasting for more than 2 hours. Stable wavelength output was observed during these repeated measurements, and no obvious degradation or irreversible wavelength drift was found. However, this method also has practical limitations. Although the pre-pump duration is short, the high peak voltage produces a large transient electrical power and a steep local temperature rise in the microheater. For the heater resistance of approximately 900 Ω used in this work, a 42 V, 10 μs pre-pump pulse corresponds to a peak electrical power of about 1.96 W and a pulse energy of about 19.6 μJ. Under this condition, wavelength switching over approximately 100 nm could be achieved. To further reduce the switching time, we also tested shorter pre-pump pulses. However, when the pulse width was reduced to 5 μs at the same amplitude of 42 V, the achievable wavelength shift decreased to only about 30 nm, indicating that a higher pulse amplitude is required to maintain a wide tuning range at shorter pulse widths. For even shorter pulses, the limitation becomes more pronounced. With a 1 μs pre-pump pulse, a 70 V amplitude, corresponding to a peak electrical power of about 5.44 W, produced a wavelength shift of less than 5 nm, while increasing the amplitude to 80 V, corresponding to a peak electrical power of about 7.11 W, caused irreversible damage to the microheater. Therefore, although reducing the pulse width decreases the pulse energy, a much higher peak power is required to achieve a comparable temperature rise within a shorter time, which increases the risk of thermal or electrical damage. These results indicate that the proposed high-power pulsed thermal tuning method involves a trade-off among switching speed, tuning range, and device reliability. Nevertheless, for applications requiring a smaller wavelength tuning range, the switching time can be further reduced by optimizing the pulse amplitude and width within the safe operating region.

In conclusion, we demonstrated a hybrid-integrated $Si_3N_4$ ECL incorporating two MRRs configured as a Vernier filter, allowing for wide-range static wavelength tuning with stable single-mode operation. Thermal actuation of the on-chip microheaters enabled laser tuning from 1486 to 1614 nm, while maintaining an output power >10 mW and a side mode suppression ratio of up to 60 dB. Two-stepped pulse thermal pumping was introduced to exploit the fast, quasilinear heating transient, overcoming the intrinsically slow TO tuning response. A 101 nm red-shift was achieved with a quasilinear tuning rate of up to 8.91 nm/μs. For blue-shift tuning, a 104 nm change was achieved at up to 9.68 nm/μs. A simple first-order thermal fit displayed good agreement with the measurements, indicating that the TO response was governed by a simple mechanism and was easy to control. This simplicity could facilitate systematic optimization through structural improvements and waveform control to further reduce the switching time. Thus, the proposed structure-agnostic pumping scheme offers a practical route to fast and wide-range wavelength switching in TO-tuned ECLs, extending their applicability to frequency-agile systems, such as wavelength-division multiplexed transceivers and LiDAR systems.